\documentclass[aps,showpacs,prd,twocolumn]{revtex4}%
\usepackage{graphicx}
\usepackage{amssymb}
\usepackage{amsmath}
\usepackage{color}
\usepackage{float}
\usepackage{ulem}
\usepackage{accents}
\usepackage{graphicx}
\usepackage{graphicx}
\usepackage{amsfonts}
\usepackage[colorlinks=true,pdfstartview=FitV,linkcolor=blue,citecolor=blue,urlcolor=blue,breaklinks=true]%
{hyperref}
\usepackage{booktabs,lipsum}%
\providecommand{\U}[1]{\protect\rule{.1in}{.1in}}
\begin{document}
\title{\textbf{Constraining EDM and MDM lepton dimension five interactions in the
electroweak sector}}
\author{Jonas B. Araujo}
\email{jonas.araujo88@gmail.com}
\author{Victor E. Mouchrek-Santos}
\email{victor\_mouchrek@hotmail.com}
\author{Frederico E. P. dos Santos}
\email{frederico.santos@ufma.br}
\author{Pedro D. S. Silva}
\email{pdiego.10@hotmail.com}
\author{Manoel M. Ferreira Jr}
\email{manojr.ufma@gmail.com}
\affiliation{Departamento de F\'{\i}sica, Universidade Federal do Maranh\~{a}o, Campus
Universit\'{a}rio do Bacanga, S\~{a}o Lu\'{\i}s - MA, 65080-805 - Brazil}
\affiliation{Coordena\c{c}\~{a}o do Curso Interdisciplinar em Ci\^{e}ncia e Tecnologia,
Universidade Federal do Maranh\~{a}o, Campus Universit\'{a}rio do Bacanga,
65080-805, S\~{a}o Lu\'{\i}s, Maranh\~{a}o, Brazil.}

\begin{abstract}
We investigate dimension five Lorentz-violating nonminimal interactions in the
electroweak sector, in connection with the possible generation of electric
dipole moment (EDM), weak electric dipole moment (WEDM), magnetic dipole
moment (MDM) and weak magnetic dipole moment (WMDM) for leptons. These
couplings are composed of the physical fields and LV tensors of ranks ranging
from 1 to 4. The CPT-odd couplings do no generate EDM\ behavior and do not
provide the correct MDM\ signature, while the CPT-even ones yield EDM and
MDM\ behavior, being subject to improved constraining. Tau lepton experimental
data is used to constrain the WEDM\ and WMDM couplings to the level of
$10^{-4}\left(  \text{GeV}\right)  ^{-1},$ whereas electron MDM and EDM data
is employed to improve constraints to the level of $10^{-17}\left(
\text{GeV}\right)  ^{-1}$ and $10^{-11}\left(  \text{GeV}\right)  ^{-1},$ respectively.

\end{abstract}

\maketitle

\section{Introduction}

Electric dipole moment (EDM) physics is a broad field of investigation
\cite{EDM1,EDM3,Yamanaka,LeptonEDM} deeply connected with precise experiments
and physics beyond the Standard Model (SM) \cite{Pospelov2005}. EDM has as
signature the violation of parity ($P)$ and time reversal ($T)$ symmetries,
while preserving charge conjugation ($C)$ and the $CPT$ symmetry. In the
relativistic context, the electric dipole moment, $\mathbf{d}%
=g(q/2m)\mathbf{S,}$ yields the interaction $d(\boldsymbol{\Sigma}%
\cdot\mathbf{E}),$ with $d$ being the EDM modulus, $\mathbf{E}$ being the
electric field, and $\boldsymbol{\Sigma}$ being the Dirac spin operator. The
EDM Lagrangian is represented by the dimension five term $d(\bar{\psi}%
i\sigma_{\mu\nu}\gamma_{5}F^{\mu\nu}\psi)$, where $\psi$ is a Dirac spinor. It
is important to mention that the EDM structure is only generated by
radioactive corrections at four-loop order \cite{LeptonEDM,SMEDM1}, so that
its magnitude is of about $d_{e}\simeq10^{-38}\,e\cdot cm$ in the SM
framework. Analogously, the magnetic dipole moment, $\mathbf{\mu
}=g(q/2m)\mathbf{S,}$ provides the relativistic magnetic interaction,
$\mu(\boldsymbol{\Sigma}\cdot\mathbf{B})$ \cite{SMEDM1,Muon1}, ${\text{ whose
Lagrangian representation, }\mu(\bar{\psi}\sigma_{\mu\nu}F^{\mu\nu}\psi),}${
appears in the }SM framework at 1-loop order.

Each order of magnitude improvement in the EDM\ experiments leads to strong
phenomenological consequences on a diversity of CP-violating theories. EDM
measurements have been progressively improved \cite{Measure1}, reaching the
level of $10^{-29}\,e\cdot{\text{cm}}$ for the electron EDM
\cite{Baron,ACME2018}, and $10^{-30}\,e\cdot{\text{cm}}$ {for the }%
$^{199}\text{Hg}$ {nuclear EDM \cite{Heckel}}.\ The gap of seven orders of
magnitude still remaining between the experimental data and the theoretical
evaluations for the electron EDM allows for new CP mechanisms, besides the
usual $CP$ violation sources already embedded in the SM. These sources may be
relevant for explaining the observed baryon asymmetry of the universe, an
issue possibly connected with axions and the strong CP problem \cite{Axion}.
EDM physics may also be related to Lorentz-violating theories, investigated in
the broader framework of the Standard Model extension (SME), developed by
Colladay and Kostelecky \cite{Colladay}. The SME incorporates dimension four
and dimension three LV terms in all sectors of the Standard Model, including
fermions \cite{fermion,fermion2,fermion3}, photons
\cite{KM1,photons1,photons2,photons3}, photon-fermion interactions
\cite{Vertex1,Vertex2}, and electroweak (EW) processes \cite{EW1,EW2,EW3}.
Beyond the minimal SME, there are nonminimal extensions encompassing couplings
with higher-order derivatives \cite{NMSME} and higher-dimensional operators
\cite{Reyes,HD,Ding}.

Lorentz violation can work as a source of CP violation and EDM generation via
radiative corrections \cite{Haghig}, or even at tree level via dimension five
nonminimal (NM) couplings \cite{Pospelov2008,Jonas1}. Dimension-five
nonminimal couplings have been proposed as nonusual QED interactions between
fermions and photons, yielding EDM Lagrangians pieces as $\lambda\bar{\psi
}(K_{F})_{\mu\nu\alpha\beta}\Gamma^{\mu\nu}F^{\alpha\beta}\psi,$ $\lambda
_{1}\bar{\psi}T_{\mu\alpha}F^{\alpha}{}_{\nu}\Gamma^{\mu\nu}\psi,$ where
$(K_{F})_{\mu\nu\alpha\beta}$\ and $T_{\mu\alpha}$ are $CPT$-even LV tensors,
with $\Gamma^{\mu\nu}=\sigma^{\mu\nu}$ or $\sigma^{\mu\nu}\gamma_{5}$
\cite{Jonas1}. Electron EDM experimental data has yielded upper bounds as
tight as $10^{-25}\,(\mbox{eV})^{-1}$ on the magnitude of these couplings.
Considering the Schiff screening theorem \cite{Schiff}, anisotropic
electrostatic interactions were taken into account in order to engender LV
corrections on the nuclear EDM and Schiff moment \cite{Jonas2}. Recently,
general dimension six nonminimal fermion-fermion couplings were proposed
\cite{Kostelecky6} and constrained at the level of $10^{-15}$ $(\text{GeV}%
)^{-2}$ by EDM data \cite{Jonas3}, considering these couplings as
electron-nucleon P-odd and T-odd atomic interactions. LV contributions to MDM
physics were also examined \cite{Stadnik,Gomes}, being constrained by precise
experimental data \cite{Gabrielse}.

If the Standard Model is addressed as a low-energy effective theory, it
becomes worthy to consider higher dimensional terms in the Lagrangian.
Extensions of the electroweak model containing higher dimension terms (mainly
dimension six) have been analyzed as effective theories since the eighties
\cite{Buch}. Lists of dimension six EW and strong couplings have been
presented and updated \cite{Grza}, so as to involve top quark physics and
interactions with the Higgs \cite{TopHiggs}. CP-violating couplings in the
Higgs sector, which comprise CP-violating interactions to quarks and tau
lepton, are also represented by dimension six operators. Such couplings can
generate EDM, providing an effective route of constraining \cite{EWEDM}. Some
of the best bounds on the anomalous CP-violating Higgs interactions come from
EDM measurements. A plethora of dimension six terms yielding electroweak
baryogenesis and CP violation has been considered in connection with the
baryon asymmetry of the universe \cite{BAU}. The role of EDM physics in
electroweak interactions and electroweak baryogenesis has been a topical issue
in the latest years \cite{BAU1,Balazs}.

Dimension five nonminimal couplings in the Glashow-Salam-Weinberg (GSW)
electroweak model have also been proposed in connection with CPT and Lorentz
symmetry violation \cite{Victor,Malta}. Such couplings have been constrained
by weak decay data at the level of $10^{-5}\,(\mbox{GeV})^{-1}$. The
repercussions of MDM and EDM physics on such nonminimal couplings has not been
examined yet, and can be used to improve constraining on these
couplings.\textbf{ }In the electroweak sector, the weak magnetic moment (WMDM)
and weak electric dipole interaction (WEDM) involve interaction with the Z
boson field, being given by the effective Lagrangian \cite{Pich,Blinov}:%
\begin{equation}
\mathcal{L}_{EW}=\frac{1}{\sin2\theta}\bar{\psi}\left[  \alpha_{w}\frac
{e}{2m_{l}}\sigma^{\mu\nu}Z_{\mu\nu}+id_{w}\sigma^{\mu\nu}\gamma_{5}Z_{\mu\nu
}\right]  \psi, \label{EDMLAG1}%
\end{equation}
where $\alpha_{w}$ and $d_{w}$ represent the WMDM\ and WEDM magnitudes,
$\theta$ is the Weinberg angle and $Z_{\mu\nu}$ is the $U(1)$ boson field
strength. Experimental limits for tau lepton WMDM and WEDM are presented in
Ref. \cite{Pich}: $\alpha_{w}<1\times10^{-3}$ and $d_{\tau}^{w}<10^{-17}e\cdot
cm.$

In this work, we analyze a few dimension five LV couplings in the GSW
electroweak model concerning the possibility of generating EDM, WEDM, MDM,
WMDM for leptons. While we propose CPT-odd and CPT-even couplings, only the
latter ones generate EDM or MDM behavior. Using tau WEDM and WMDM experimental
data, some couplings are constrained to the level of $10^{-5}\left(
\text{GeV}\right)  ^{-1}$, while the electron\ EDM and MDM\textbf{ }yield
upper bounds to the level of $10^{-17}$ $(\text{GeV})^{-1}$ and $10^{-11}$
$(\text{GeV})^{-1}$, respectively.

\section{The Glashow-Salam-Weinberg electroweak model}

In the GSW model, the left-handed leptons are disposed in isodublets
($T=\frac{1}{2},T_{3}=\pm\frac{1}{2}$), while the the right-handed leptons are
represented by isosinglets ($T=0)$ under the $SU\left(  2\right)  $ group,%
\begin{align}
L_{l}  &  =%
\begin{bmatrix}
\psi_{\nu_{l}}\\
\psi_{l}%
\end{bmatrix}
_{L}=\frac{1-\gamma_{5}}{2}%
\begin{bmatrix}
\psi_{\nu_{l}}\\
\psi_{l}%
\end{bmatrix}
,\label{LGSW1}\\
R_{l}  &  =\left(  \psi_{l}\right)  _{R}=\left(  \frac{1+\gamma_{5}}%
{2}\right)  \psi_{l}, \label{LGSW2}%
\end{align}
with the generators, $\mathbf{T=}\left(  T_{1},T_{2},T_{3}\right)  $,
fulfilling the relation, $\left[  T_{i},T_{j}\right]  =i\varepsilon_{ijk}%
T_{k}.$ The GSW Lagrangian is
\begin{equation}
\mathcal{L}=\bar{L}_{l}\gamma^{\mu}iD_{\mu}L_{l}+\bar{R}_{l}\gamma^{\mu
}iD_{\mu}R_{l}-\frac{1}{4}\mathbf{W}_{\mu\nu}\cdot\mathbf{W}^{\mu\nu}-\frac
{1}{4}B_{\mu\nu}B^{\mu\nu}, \label{LGSW.14}%
\end{equation}
where the field strengths for the $U(1)$ and $SU(2)$ gauge fields, $B_{\mu}$
and $\mathbf{W}_{\mu},$ are $B_{\mu\nu}=\partial_{\mu}B_{\nu}-\partial_{\nu
}B_{\mu},$ and
\begin{equation}
\mathbf{W}_{\mu\nu}=\partial_{\mu}\mathbf{W}_{\nu}-\partial_{\nu}%
\mathbf{W}_{\mu}+g\left(  \mathbf{W}_{\mu}\times\mathbf{W}_{\nu}\right)  .
\label{LGSW.10}%
\end{equation}
Knowing that the $U(1)$ field is a combination of the electromagnetic and the
boson Z field, $B_{\mu}=\cos\theta A_{\mu}-\sin\theta Z_{\mu},$ one has
\begin{equation}
B_{\mu\nu}=\left(  \cos\theta\right)  F_{\mu\nu}-\left(  \sin\theta\right)
Z_{\mu\nu}, \label{B_FS}%
\end{equation}
with $F_{\mu\nu}=\partial_{\mu}A_{\nu}-\partial_{\nu}A_{\mu}$, and $Z_{\mu\nu
}=\partial_{\mu}Z_{\nu}-\partial_{\nu}Z_{\mu}.$The usual covariant derivative
is%
\begin{equation}
D_{\mu}=\partial_{\mu}-ig\left(  \mathbf{T\cdot W}\right)  _{\mu}%
-i\frac{g^{\prime}}{2}YB_{\mu}. \label{LGSW.7}%
\end{equation}
where $Y$ is the $U(1)$ generator. We have $Y_{L}=-1$ for $\left(  e_{L}%
,\mu_{L},\tau_{L}\right)  $ and $Y_{R}=-2$ for $\left(  e_{R},\mu_{R},\tau
_{R}\right)  $. Replacing the covariant derivative in the Lagrangian
(\ref{LGSW.14}), we obtain%
\begin{equation}
\mathcal{L}=i\bar{L}_{l}\gamma^{\mu}\partial_{\mu}L_{l}+i\bar{R}_{l}%
\gamma^{\mu}\partial_{\mu}R_{l}+L_{int}^{\left(  l\right)  },
\end{equation}
with the interaction piece being
\begin{equation}
\mathcal{L}_{int}^{\left(  l\right)  }=g\left(  \bar{L}_{l}\gamma^{\mu
}\mathbf{T}L_{l}\right)  \cdot\mathbf{W}_{\mu}-\left[  \frac{g^{\prime}}%
{2}\left(  \bar{L}_{l}\gamma^{\mu}L_{l}\right)  +g^{\prime}\left(  \bar{R}%
_{l}\gamma^{\mu}R_{l}\right)  \right]  B_{\mu}. \label{Lint1}%
\end{equation}

\section{CPT-odd dimension five nonminimal LV electroweak coupling}

We investigate some CPT-odd nonminimal couplings. They do not generate EDM nor
possess the correct MDM signature under CPT operators. This can be argued by
analyzing rank-1 or rank-3 nonminimal couplings, which are the simplest ones
to be proposed.

\subsection{Rank-1 CPT-odd NMC}

Rank-1 CPT-odd and dimension five nonminimal coupling in the EW sector were
proposed in Ref. \cite{Victor}, as
\begin{equation}
\mathcal{L}_{Int}=g_{1}^{\prime}\left(  \bar{L}_{l}\gamma^{\mu}B_{\mu\nu
}C^{\nu}L_{l}\right)  +2g_{1}^{\prime}\left(  \bar{R}_{l}\gamma^{\mu}B_{\mu
\nu}C^{\nu}R_{l}\right)  , \label{NMCL1}%
\end{equation}
where $C^{\nu}$ is a fixed LV background, and $Y=-1$ and $Y=-2$ for
left-handed and right-handed fermions, respectively. Using Eq. (\ref{LGSW1})
and (\ref{LGSW2}), we obtain the Lagrangian:%
\begin{align}
\mathcal{L}_{int}  &  =\frac{1}{2}g_{1}^{\prime}\bar{\psi}_{\nu_{l}}%
\gamma^{\mu}\left(  1-\gamma_{5}\right)  \psi_{\nu_{l}}B_{\mu\nu}C^{\nu
}\nonumber\\
&  +\frac{1}{2}g_{1}^{\prime}\bar{\psi}_{l}\gamma^{\mu}\left(  3+\gamma
_{5}\right)  \psi_{l}B_{\mu\nu}C^{\nu}. \label{NMCL2}%
\end{align}
Here, it is important to note that the $\gamma_{5}$ operator changes the
behavior of the coupling in relation to the $C$, $P$ and $T$ operators. Thus,
it is suitable to rewrite Lagrangian (\ref{NMCL2}) in terms of two vector
backgrounds, $C^{\nu}$ and $C_{A}^{\nu}$:%
\begin{align}
\mathcal{L}_{(1)}^{(odd)}  &  =\frac{1}{2}g_{1}^{\prime}\bar{\psi}_{\nu_{l}%
}\gamma^{\mu}\psi_{\nu_{l}}B_{\mu\nu}C^{\nu}-\frac{1}{2}g_{1}^{\prime}%
\bar{\psi}_{\nu_{l}}\gamma^{\mu}\gamma_{5}\psi_{\nu_{l}}B_{\mu\nu}C_{A}^{\nu
}\nonumber\\
&  +\frac{3}{2}g_{1}^{\prime}\bar{\psi}_{l}\gamma^{\mu}\psi_{l}B_{\mu\nu
}C^{\nu}+\frac{1}{2}g_{1}^{\prime}\bar{\psi}_{l}\gamma^{\mu}\gamma_{5}\psi
_{l}B_{\mu\nu}C_{A}^{\nu}. \label{NMCL3}%
\end{align}
For purpose of better investigation, one explicitly analyzes the lepton ($l$)
content of Eq. (\ref{NMCL3}), writing
\begin{align}
\mathcal{L}_{(1)l}^{(odd)}  &  =\frac{1}{2}g_{1}^{\prime}\left[  3\bar{\psi
}_{l}\gamma^{0}B_{0i}C^{i}\psi_{l}+\bar{\psi}_{l}\gamma^{0}\gamma_{5}%
B_{0i}C_{A}^{i}\psi_{l}\right. \nonumber\\
&  +3\bar{\psi}_{l}\gamma^{i}B_{ij}C^{j}\psi_{l}+\bar{\psi}_{l}\gamma
^{i}\gamma_{5}B_{ij}C_{A}^{j}\psi_{l}\nonumber\\
&  \left.  +3\bar{\psi}_{l}\gamma^{i}B_{i0}C^{0}\psi_{l}+\bar{\psi}_{l}%
\gamma^{i}\gamma_{5}B_{i0}C_{A}^{0}\psi_{l}\right]  . \label{NMCL4}%
\end{align}
\begin{table}[t]
\centering%
\begin{tabular}
[c]{|c|c|c|c|c|}\hline
Coupling & $g_{1}^{\prime}C^{0}$ & $g_{1}^{\prime}C^{i}$ & $g_{1}^{\prime
}C_{A}^{0}$ & $g_{1}^{\prime}C_{A}^{i}$\\\hline
P & $+$ & $-$ & $-$ & $+$\\\hline
C & $-$ & $+$ & $-$ & $-$\\\hline
T & $+$ & $+$ & $-$ & $+$\\\hline
\end{tabular}
\caption{Classification under $C,P,T$ \ for the $CPT$-odd nonminimal couplings
of Lagrangian (\ref{NMCL3}).}%
\label{Tab1}%
\end{table}Lepton Lagrangian term $\bar{\psi}_{l}\gamma^{i}\gamma_{5}\psi
_{l}B_{i0}C_{A}^{0}$ is the unique one compatible with the EDM signature, as
shown in Table (\ref{Tab1}), since it contains the pieces
\begin{equation}
\left(  \bar{\psi}_{l}\gamma^{0}\Sigma^{i}\psi_{l}\right)  B_{0i}C_{A}^{0},
\label{EDMcptodd1}%
\end{equation}
where $B_{0i}=E^{i}+\tilde{E}^{i}$ could yield electric and electroweak EDM,
with $\tilde{E}^{i}$ representing the weak electric field, as shown in
(\ref{Z0ib}). This same analysis holds equally for the neutrino terms in
Lagrangian (\ref{NMCL2}), where the EDM-like term is $\left(  \bar{\psi}%
_{\nu_{l}}\gamma^{0}\Sigma^{i}\psi_{\nu_{l}}\right)  B_{0i}C^{0}.$ The
presence of the $\gamma^{0}$, however, prevents the EDM behavior, since it
avoids the resemblance to the EDM Lagrangian $\left(  \bar{\psi}_{l}\Sigma
^{j}E^{j}{}\psi_{l}\right)  $. The $\gamma^{0}$ factor disappears in the
corresponding Hamiltonian form, yielding the\ relativistic interaction pieces,
$\Sigma^{j}E^{j},\Sigma^{j}\tilde{E}^{j},$ which are undetectable (at first
order) in accordance with the Schiff%
\'{}%
s theorem \cite{LeptonEDM,Schiff}. This occurs with the Hamiltonian
interactions stemming from Eq. (\ref{EDMcptodd1}),
\begin{equation}
\psi_{l}^{\dagger}\Sigma^{i}E^{i}C^{0}\psi_{l},\text{ \ \ }\psi_{l}^{\dagger
}\Sigma^{i}Z_{0i}C^{0}\psi_{l},
\end{equation}
implying absence of EDM physics. Here, we have used%
\begin{align}
\sigma^{0j}  &  =i\alpha^{j},\text{ \ }\sigma^{ij}=\epsilon_{ijk}\Sigma
^{k},\label{sigma0i}\\
\alpha^{j}{}\gamma_{5}  &  =\Sigma^{j}~,~\text{\ }\alpha^{j}=\ \Sigma
^{j}\gamma_{5},\label{alphajgamma5}\\
F_{0j}  &  =E^{j},\text{ }F_{mn}=\epsilon_{mnp}B^{p},\label{F0ib}\\
Z_{0j}  &  =\tilde{E}^{j},\text{ }Z_{mn}=\epsilon_{mnp}\tilde{B}^{p}.
\label{Z0ib}%
\end{align}
We can also investigate MDM behavior for leptons and neutrinos. The closest
MDM term in Lagrangian (\ref{NMCL3}) is%
\begin{equation}
\text{ }\bar{\psi}_{l}\gamma^{i}\gamma_{5}B_{ij}C_{A}^{j}\psi_{l},
\end{equation}
where $B_{ij}=F_{ij}+Z_{ij}.$ Considering the definitions (\ref{F0ib}) and
(\ref{Z0ib})$,$ this term involves the usual ($B^{k})$ and weak ($\tilde
{B}^{k})$ magnetic fields,%
\begin{equation}
\bar{\psi}_{l}\gamma^{0}\left(  \mathcal{C}_{A}\right)  _{ik}\Sigma^{i}%
B^{k}\psi_{l},\text{ \ \ }\bar{\psi}_{l}\gamma^{0}\left(  \mathcal{C}%
_{A}\right)  _{ik}\Sigma^{i}\tilde{B}^{k}\psi_{l}, \label{cptod1mag}%
\end{equation}
in a non conventional way (it couples the spin to a \textquotedblleft rotated"
LV background structure, $\left(  \mathcal{C}_{A}\right)  _{ik}=\epsilon
_{jik}C_{A}^{j}).$ As shown in Table (\ref{Tab1}), the coefficient $C_{A}^{j}$
has not the exact signature of a MDM interaction, due the presence of
$\gamma^{0}$. Thus, it does not generate MDM or WMDM. For experimental
purposes, as this tensor background has no diagonal components, $\mathcal{C}%
_{ii}=0,$ the contributions (\ref{cptod1mag}) could only be probed with a
magnetic field orthogonal to the spin, as discussed in Refs.
\cite{Jonas1,Jonas2}. The same conclusions hold for the neutrino counterparts.

\subsection{Rank-3 CPT-odd NMC}

We now examine a rank$-3$ dimension five nonminimal coupling in Lagrangian of
the GSW model. It can be written as%
\begin{align}
\mathcal{L}_{(3)}^{\left(  odd\right)  }  &  =-g_{2}^{\prime}Y_{L}\bar{L}%
_{l}\left(  \gamma^{\mu}B^{\alpha\beta}H_{\mu\alpha\beta}\right)
L_{l}\nonumber\\
&  -g_{2}^{\prime}Y_{R}\bar{R}_{l}\left(  \gamma^{\mu}B^{\alpha\beta}%
H_{\mu\alpha\beta}\right)  R_{l}, \label{R3NM0.1}%
\end{align}
where $H_{\mu\alpha\beta}$ is the LV background tensor, with the supposed
symmetry $H_{\mu\alpha\beta}=-H_{\mu\beta\alpha},$ and $Y_{L}=-1,Y_{R}=-2$, so
that%
\begin{equation}
\mathcal{L}_{(3)}^{\left(  odd\right)  }=g_{2}^{\prime}\bar{L}_{l}\left(
\gamma^{\mu}B^{\alpha\beta}H_{\mu\alpha\beta}\right)  L_{l}+g_{2}^{\prime}%
\bar{R}_{l}\left(  \gamma^{\mu}B^{\alpha\beta}H_{\mu\alpha\beta}\right)
R_{l}. \label{sigma8}%
\end{equation}
This EW Lagrangian can be written in terms of the lepton and neutrino pieces,
$\mathcal{L}_{(3)}^{\left(  odd\right)  }=\mathcal{L}_{(3)l}^{\left(
odd\right)  }+\mathcal{L}_{(3)\nu}^{\left(  odd\right)  },$ given as
\begin{align}
\mathcal{L}_{(3)l}^{\left(  odd\right)  }  &  =\frac{g_{2}^{\prime}}{2}%
\bar{\psi}_{l}\left[  3\gamma^{\mu}H_{\mu\alpha\beta}B^{\alpha\beta}%
+\gamma^{\mu}\gamma_{5}B^{\alpha\beta}\left(  H_{A}\right)  _{\mu\alpha\beta
}\right]  \psi_{l},\label{sigma9.3}\\
\mathcal{L}_{(3)\nu}^{\left(  odd\right)  }  &  =\frac{g_{2}^{\prime}}{2}%
\bar{\psi}_{\nu_{l}}\left[  \gamma^{\mu}H_{\mu\alpha\beta}B^{\alpha\beta
}-\gamma^{\mu}\gamma_{5}B^{\alpha\beta}\left(  H_{A}\right)  _{\mu\alpha\beta
}\right]  \psi_{\nu_{l}},
\end{align}
where we have introduced the rank-3 background $\left(  H_{A}\right)
_{\mu\alpha\beta}$ for the coupling involving $\gamma_{5},$ as we have done in
Eq. (\ref{NMCL3}). In order to verify the possibility of EDM generation for
leptons, we investigate the tensor structure of lepton Lagrangian
(\ref{sigma9.3}) that can expressed as%
\begin{gather}
\mathcal{L}_{(3)l}^{\left(  odd\right)  }=g_{2}^{\prime}\left[  \bar{\psi}%
_{l}\gamma^{0}B^{0i}H_{00i}\psi_{l}+\bar{\psi}_{l}\gamma^{0}B^{ij}H_{0ij}%
\psi_{l}\right. \nonumber\\
+\bar{\psi}_{l}\gamma^{i}B^{0j}H_{i0j}\psi_{l}+\bar{\psi}_{l}\gamma^{i}%
B^{jk}H_{ijk}\psi_{l}\label{R3NM6.7}\\
+\bar{\psi}_{l}\gamma^{0}\gamma_{5}B^{0i}\left(  H_{A}\right)  _{00i}\psi
_{l}+\bar{\psi}_{l}\gamma^{0}\gamma_{5}B^{ij}\left(  H_{A}\right)  _{0ij}%
\psi_{l}\nonumber\\
\left.  +\bar{\psi}_{l}\gamma^{i}\gamma_{5}B^{0j}\left(  H_{A}\right)
_{i0j}\psi_{l}+\bar{\psi}_{l}\gamma^{i}\gamma_{5}B^{jk}\left(  H_{A}\right)
_{ijk}\psi_{l}\right]  .\nonumber
\end{gather}
In Eq. (\ref{R3NM6.7}), we see that the term, $\bar{\psi}_{l}\gamma^{i}%
\gamma_{5}B^{0j}\left(  H_{A}\right)  _{i0j}\psi_{l},$ is the unique that has
EDM signature, as shown in Table (\ref{Tab1b}). This piece can be written as%
\begin{equation}
\bar{\psi}_{l}\gamma^{0}\Sigma^{i}B_{0j}{}\left(  H_{A}\right)  _{i0j}\psi
_{l},
\end{equation}
in which $B_{0j}$ contains the electric and weak electric counterparts.
Analogously to the rank$-1$ CPT-odd NM coupling, the presence of the
$\gamma^{0}$ avoids the EDM behavior. \begin{table}[t]
\centering%
\begin{tabular}
[c]{|c|c|c|c|c|}\hline
Coupling & $g_{2}^{\prime}H_{00i}$ & $g_{2}^{\prime}H_{0ij}$ & $g_{2}^{\prime
}H_{i0j}$ & $g_{2}^{\prime}H_{ijk}$\\\hline
P & $-$ & $+$ & $+$ & $-$\\\hline
C & $+$ & $+$ & $-$ & $+$\\\hline
T & $+$ & $-$ & $+$ & $+$\\\hline
Coupling & $g_{2}^{\prime}\left(  H_{A}\right)  _{00i}$ & $g_{2}^{\prime
}\left(  H_{A}\right)  _{0ij}$ & $g_{2}^{\prime}\left(  H_{A}\right)  _{i0j}$
& $g_{2}^{\prime}\left(  H_{A}\right)  _{ijk}$\\\hline
P & $+$ & $+$ & $-$ & $+$\\\hline
C & $-$ & $-$ & $-$ & $-$\\\hline
T & $+$ & $+$ & $-$ & $+$\\\hline
\end{tabular}
\caption{Classification under $C,P,T$ \ for the $CPT$-odd rank-3 nonminimal
couplings of Lagrangian (\ref{R3NM6.7}).}%
\label{Tab1b}%
\end{table}As it occurs for the rank-1 case, the Table (\ref{Tab1b}) shows
that the couplings of Lagrangian (\ref{R3NM6.7}) do not possess MDM behavior.

There are another possibilities of writing (hermitian) rank-3 nonminimal
couplings. An example is%
\begin{align}
\mathcal{\tilde{L}}_{(3)l}^{\left(  odd\right)  }  &  =g_{3}^{\prime}\bar
{\psi}_{l}(\gamma^{\alpha}B{}^{\beta\nu}-\gamma^{\beta}B{}^{\alpha\nu}%
)\psi_{l}\bar{H}_{\nu\alpha\beta}\nonumber\\
&  +g_{3}^{\prime}\bar{\psi}_{l}(\gamma^{\alpha}B{}^{\beta\nu}-\gamma^{\beta
}B{}^{\alpha\nu})\gamma_{5}\psi_{l}\left(  \bar{H}_{A}\right)  _{\nu
\alpha\beta}.
\end{align}
These couplings do not generate EDM behavior, do not posses MDM correct
signature, and will be no longer examined.

\section{CPT-even dimension five nonminimal LV electroweak couplings}

In this section, we analyze CPT-even dimension five nonminimal couplings
composed of rank-2 and rank-4 tensors, which generate EDM and MDM behavior.

\subsection{Rank$-2$ nonminimal coupling}

The EDM Lagrangian terms should have the form presented in Eq. (\ref{EDMLAG1}%
). Initially, the idea could be to propose a form written in terms of a
covariant derivative into the interaction Lagrangian (\ref{Lint1}). In the
hermitian form, we first propose a non axial \textbf{(}without\textbf{
}$\gamma_{5}$) modified covariant derivative,%
\begin{equation}
\mathcal{D}_{\mu}=D_{\mu}-\frac{i}{2}\lambda_{1}\left(  T_{\mu\nu}B^{\nu\beta
}-T^{\beta\nu}B_{\nu\mu}\right)  \gamma_{\beta}, \label{NMCeven1}%
\end{equation}
based on the pattern first analyzed in Ref. \cite{Jonas1}. Replacing this
covariant derivative in the EW quiral Lagrangian structure for left-handed
leptons, $\bar{L}_{l}\gamma^{\mu}iD_{\mu}L_{l},$ we obtain%
\begin{equation}
\mathcal{L}=\bar{L}_{l}\gamma^{\mu}i\left[  -\frac{i}{2}\lambda_{1}\left(
T_{\mu\nu}B^{\nu\beta}-T^{\beta\nu}B_{\nu\mu}\right)  \gamma_{\beta}\right]
L_{l}.
\end{equation}
Using the identity, $\gamma^{\mu}\gamma_{\beta}=\left(  \delta^{\mu}{}_{\beta
}-i\sigma^{\mu}{}_{\beta}\right)  ,$ it becomes%
\begin{equation}
\mathcal{L}=-i\lambda_{1}\bar{L}_{l}\left[  \sigma^{\mu\beta}{}\left(
T_{\mu\nu}B_{\text{ }\beta}^{\nu}\right)  \right]  L_{l}, \label{NMCeven2L}%
\end{equation}
where it was neglected a term of the form $\bar{L}_{l}[\left(  T_{\beta\nu
}B^{\nu\beta}\right)  ]L_{l},$ since it does not contain any gamma matrices
nor spin components. Now it is necessary to remark that this nonminimal
coupling is not properly communicated to the Lagrangian pieces of leptons and
neutrinos. Indeed, we notice that%
\begin{equation}
\left(  \frac{1\pm\gamma_{5}}{2}\right)  X\left(  \frac{1\mp\gamma_{5}}%
{2}\right)  =0, \label{G5X}%
\end{equation}
if the operator $X$ contains an even number of gamma matrices, which includes
$X=\sigma^{\mu\beta}$ as a special case. Otherwise, if the operator $X$
possesses an odd number of gamma matrices, the quantity in Eq. (\ref{G5X}) is
not null, in principle. Thus, Lagrangian (\ref{NMCeven2L}) yields a null
contribution; the same holds for the right-handed fermions:
\begin{equation}
\bar{L}_{l}\left[  \sigma^{\mu\beta}{}\left(  T_{\mu\nu}B_{\text{ \ }\beta
}^{\nu}\right)  \right]  L_{l}=\bar{R}_{l}\left[  \sigma^{\mu\beta}{}\left(
T_{\mu\nu}B_{\text{ \ }\beta}^{\nu}\right)  \right]  R_{l}=0.
\end{equation}
In order to circumvent this difficulty, we can propose $U(1)$ CPT-even NM
couplings directly on the\textbf{ }neutrino and lepton Lagrangian spinors:%
\begin{align}
\mathcal{L}_{(2)l}^{(even)}  &  =\lambda_{l}\bar{\psi}_{_{l}}\left[
\sigma^{\mu\beta}T_{\mu\nu}B_{\text{ \ }\beta}^{\nu}-i\sigma^{\mu\beta}%
\gamma_{5}R_{\mu\nu}B_{\text{ \ }\beta}^{\nu}\right]  \psi_{_{l}%
},\label{Lepton1}\\
\mathcal{L}_{(2)\nu}^{(even)}  &  =\lambda_{\nu_{l}}\bar{\psi}_{\nu_{l}%
}\left[  \sigma^{\mu\beta}T_{\mu\nu}B_{\text{ \ }\beta}^{\nu}-i\sigma
^{\mu\beta}\gamma_{5}R_{\mu\nu}B_{\text{ \ }\beta}^{\nu}\right]  \psi_{\nu
_{l}}. \label{Neutrino1}%
\end{align}
where the imaginary factor was introduced with the matrix $\gamma_{5}$ in
order to assure hermiticity. The leptons NM couplings in Eq. (\ref{Lepton1})
exhibit a \textquotedblleft non axial" (without $\gamma_{5})$ and an
\textquotedblleft axial" (with $\gamma_{5})$ interaction piece:%
\begin{align}
\mathcal{L}_{(2)l(T)}^{(even)}=  &  \lambda_{l}\bar{\psi}_{_{l}}\sigma
^{\mu\beta}{}\left(  T_{\mu\nu}B_{\text{ \ }\beta}^{\nu}\right)  \psi_{_{l}%
},\text{ \ }\label{Lepton1piece}\\
\mathcal{L}_{(2)l(A)}^{(even)}=  &  i\lambda_{l}\bar{\psi}_{_{l}}\left(
\sigma^{\mu\beta}\gamma_{5}R_{\mu\nu}B_{\text{ \ }\beta}^{\nu}\right)
\psi_{_{l}}, \label{Lepton2piece}%
\end{align}
where the label $(T)$ refers to the tensor $T_{\mu\nu}$ and the label $(A)$
refers to the \textquotedblleft axial" tensor $\gamma_{5}R_{\mu\nu}$ coupling.
Such couplings are represented by two distinct tensors, $T_{\mu\nu}$ and
$R_{\mu\nu}$, to stress that the interactions with and without $\gamma_{5}$
are physically different.

The lepton first piece can be explicitly written as%
\begin{gather}
\mathcal{L}_{(T)}^{l}=\cos\theta\bar{\psi}_{_{l}}\left[  i\lambda_{l}%
\alpha^{i}T_{00}E^{i}+i\lambda_{l}\epsilon_{aip}\alpha^{i}T_{0a}B^{p}\right]
\nonumber\\
-i\lambda_{l}\alpha^{i}T_{ij}E^{j}+\lambda_{l}\mathcal{T}_{jk}\Sigma^{k}%
E^{j}\nonumber\\
\left.  +\lambda_{l}T_{ii}\Sigma^{k}B^{k}-\lambda_{l}T_{ik}\Sigma^{k}%
B^{i}\right]  \psi_{_{l}}-\sin\theta\bar{\psi}_{_{l}}\left[  i\lambda
_{l}\alpha^{i}T_{00}\tilde{E}_{\text{ }}^{i}\right. \nonumber\\
+i\lambda_{l}\alpha^{i}T_{0a}\epsilon_{aik}\tilde{B}^{k}-i\lambda_{l}%
T_{ij}\alpha^{i}\tilde{E}_{\text{ }}^{j}\nonumber\\
\left.  +\lambda_{l}\mathcal{T}_{jk}\Sigma^{k}\tilde{E}_{\text{ }}^{j}%
+\lambda_{l}T_{ii}\Sigma^{k}\tilde{B}^{k}-\lambda_{l}T_{ik}\Sigma^{k}\tilde
{B}^{i}\right]  \psi_{_{l}}, \label{Lepton1_1}%
\end{gather}
where we have used the conventions (\ref{sigma0i}), (\ref{alphajgamma5}),
(\ref{F0ib}) and (\ref{Z0ib}). Such an expression provides \textquotedblleft
rotated" EDM and weak EDM contributions:%
\begin{align}
\mathcal{L}_{(T)EDM}^{l}  &  =\lambda_{l}\cos\theta\bar{\psi}_{_{l}}\left(
\mathcal{T}_{jk}\Sigma^{k}{}E^{j}\right)  \psi_{_{l}},\text{ }\\
\mathcal{L}_{(T)WEDM}^{l}  &  =\text{\ }-\lambda_{l}\sin\theta\bar{\psi}%
_{_{l}}\left(  \mathcal{T}_{jk}\Sigma^{k}{}\tilde{E}^{j}\right)  \psi_{_{l}}.
\label{LWEDM1}%
\end{align}
having as counterpart the following Hamiltonian contributions:
\begin{align}
\mathcal{H}_{EDM}^{l}  &  =-\lambda_{l}\cos\theta\psi_{l}^{\dagger}\gamma
^{0}\left(  \mathcal{T}_{jk}\Sigma^{k}E^{j}\right)  \psi_{l},\text{
}\label{EDM1}\\
\mathcal{H}_{WEDM}^{l}  &  =\text{\ }\lambda_{l}\sin\theta\psi_{l}^{\dagger
}\gamma^{0}\left(  \mathcal{T}_{jk}\Sigma^{k}\tilde{E}_{\text{ }}^{j}\right)
\psi_{l}, \label{WDM1}%
\end{align}
with the $\gamma^{0}$ factor\textbf{ }circumventing the Schiff theorem
\cite{Schiff} and assuring the effective EDM character. The EDM signature is
also revealed by the behavior of these couplings under C,P,T operations, as
shown in Table (\ref{TableCPT2}). Here, $\mathcal{T}_{jk}$ is a
\textquotedblleft rotated" background redefined as%
\begin{equation}
\mathcal{T}_{jk}=\epsilon_{ijk}{}T_{i0}, \label{RB}%
\end{equation}
that allows to write the interactions in a more direct way.

In expression (\ref{Lepton1_1}), we also identify MDM\ and weak MDM (WMDM)
interactions for leptons associated with the Lagrangian terms:%
\begin{align}
\mathcal{L}_{(T)(MDM)}^{l} &  =\lambda_{l}\left(  \cos\theta\right)  \left[
\bar{\psi}_{_{l}}\left(  {}T\Sigma^{k}B_{\text{ \ }}^{k}\right)  \psi_{_{l}%
}\right.  \nonumber\\
&  \left.  -\bar{\psi}_{_{l}}{}\left(  T_{ik}\Sigma^{k}B^{i}\right)
\psi_{_{l}}\right]  ,\\
\mathcal{L}_{(T)(WMDM)}^{l} &  =\lambda_{l}\left(  \sin\theta\right)  \left[
-\bar{\psi}_{_{l}}\left(  T\Sigma^{k}\tilde{B}_{\text{ \ }}^{k}\right)
\psi_{_{l}}\right.  \nonumber\\
&  \left.  +\bar{\psi}_{_{l}}\left(  T_{ik}\Sigma^{k}\tilde{B}^{i}\right)
\psi_{_{l}}\right]  ,\label{LWMDM}%
\end{align}
where $T=T_{ii}$ $=$Tr$(T_{ii})$ is the trace of space sector of the tensor
$T_{\mu\nu}.$ Analogously, we can perform the same analysis for the second
lepton piece (\ref{Lepton2piece}), whose tensor structure is%
\begin{gather}
\mathcal{L}_{(A)}^{l}=\cos\theta\bar{\psi}_{_{l}}\left[  \lambda_{l}%
R_{00}\Sigma^{i}E^{i}+\lambda_{l}\mathcal{R}_{ik}\Sigma^{i}B^{k}-\lambda
_{l}R_{ij}\Sigma^{i}E^{j}\right.  \nonumber\\
\left.  -i\lambda_{l}\epsilon_{ijk}\alpha^{k}R_{i0}E^{j}-i\lambda_{l}%
\alpha^{k}R_{ii}B^{k}+i\lambda_{l}\alpha^{k}R_{ik}B^{i}\right]  \bar{\psi
}_{_{l}}\nonumber\\
-\sin\theta\bar{\psi}_{_{l}}\left[  \lambda_{l}R_{00}\Sigma^{i}\tilde{E}%
^{i}+\lambda_{l}\mathcal{R}_{ik}\Sigma^{i}\tilde{B}^{k}-\lambda_{l}%
R_{ij}\Sigma^{i}\tilde{E}^{j}\right.  \nonumber\\
\left.  -i\lambda_{l}\epsilon_{ijk}\alpha^{k}R_{i0}\tilde{E}^{j}-i\lambda
_{l}\alpha^{k}R_{ii}\tilde{B}^{k}+i\lambda_{l}\alpha^{k}R_{ik}\tilde{B}%
^{i}\right]  \psi_{_{l}},\label{Lepton1_2}%
\end{gather}
where we have used the relations (\ref{sigma0i}), (\ref{alphajgamma5}),
(\ref{F0ib}), (\ref{Z0ib}), and $\mathcal{R}_{jk}=\epsilon_{ijk}{}R_{i0}.$
Such an expression provides two direct Lorentz-violating EDM contributions for
leptons:%
\begin{align}
\mathcal{L}_{(A)(EDM)}^{l} &  =\lambda_{l}\cos\theta\left[  \bar{\psi}_{_{l}%
}\left(  R_{00}\Sigma^{i}E_{\text{ }}^{i}\right)  \psi_{_{l}}\right.
\nonumber\\
&  \left.  -\bar{\psi}_{_{l}}\left(  R_{ij}\Sigma^{i}E^{j}\right)  \psi_{_{l}%
}\right]  ,
\end{align}
and two direct Lorentz-violating weak EDM\ (weak dipole moment) pieces:%
\begin{align}
\mathcal{L}_{(A)(WEDM)}^{l} &  =\lambda_{l}\sin\theta\left[  -\bar{\psi}%
_{_{l}}\left(  R_{00}\Sigma^{i}\tilde{E}^{i}\right)  \psi_{_{l}}\right.
\nonumber\\
&  \left.  +\bar{\psi}_{_{l}}\left(  R_{ij}\Sigma^{i}\tilde{E}^{j}\right)
\psi_{_{l}}\right]  .\label{LWEDM2}%
\end{align}
There are rotated lepton MDM and weak MDM contributions as well:%
\begin{align}
\mathcal{L}_{(A)(MDM)}^{l} &  =\lambda_{l}\cos\theta\bar{\psi}_{_{l}}\left(
\mathcal{R}_{ik}\Sigma^{i}B^{k}\right)  \psi_{_{l}},\\
\mathcal{L}_{(A)(WMDM)}^{l} &  =\lambda_{l}\sin\theta\bar{\psi}_{_{l}}\left(
\mathcal{R}_{ik}\Sigma^{i}\tilde{B}_{\text{ \ }}^{k}\right)  \psi_{_{l}}.
\end{align}

All these terms are shown in Table (\ref{T1a}), which contains the EDM, WEDM,
MDM and WMDM contributions to the Hamiltonian of the lepton NM coupling in Eq.
(\ref{Lepton1}).


\begin{table}[t]
\centering%
\begin{tabular}
[c]{|c|c|}\hline
EDM & WEDM\\\hline
$-\lambda_{l}\cos\theta\psi_{l}^{\dagger}\gamma^{0}\left(  R_{00}\Sigma
^{i}E_{\text{ }}^{i}\right)  \psi_{_{l}}$ & $\lambda_{l}\sin\theta\psi
_{l}^{\dagger}\gamma^{0}\left(  R_{00}\Sigma^{i}\tilde{E}_{\text{ }}%
^{i}\right)  \psi_{_{l}}$\\\hline
$-\lambda_{l}\cos\theta\psi_{l}^{\dagger}\gamma^{0}\left(  R_{jk}\Sigma
^{k}E^{j}\right)  \psi_{l}$ & $\lambda_{l}\sin\theta\psi_{l}^{\dagger}%
\gamma^{0}\left(  R_{jk}\Sigma^{k}\tilde{E}_{\text{ }}^{j}\right)  \psi_{l}%
$\\\hline
$\lambda_{l}\cos\theta\psi_{l}^{\dagger}\gamma^{0}\left(  \mathcal{T}%
_{ij}\Sigma^{i}E^{j}\right)  \psi_{_{l}}$ & $-\lambda_{l}\sin\theta\psi
_{l}^{\dagger}\gamma^{0}\left(  \mathcal{T}_{ij}\Sigma^{i}\tilde{E}_{\text{ }%
}^{j}\right)  \psi_{_{l}}$\\\hline
& \\\hline
MDM & \ WMDM\\\hline
$-\lambda_{l}\cos\theta\psi_{l}^{\dagger}\gamma^{0}\left(  T\Sigma^{k}%
B^{k}\right)  \psi_{l}$ & $\lambda_{l}\sin\theta\psi_{l}^{\dagger}\gamma
^{0}\left(  T\Sigma^{k}\tilde{B}^{k}\right)  \psi_{l}$\\\hline
$-\lambda_{l}\cos\theta\psi_{l}^{\dagger}\gamma^{0}\left(  T_{ik}\Sigma
^{i}B^{k}\right)  \psi_{_{l}}$ & $\lambda_{l}\sin\theta\psi_{l}^{\dagger
}\gamma^{0}\left(  T_{ik}\Sigma^{i}\tilde{B}_{\text{ \ }}^{k}\right)
\psi_{_{l}}$\\\hline
$\lambda_{l}\cos\theta\psi_{l}^{\dagger}\gamma^{0}\left(  \mathcal{R}%
_{ik}\Sigma^{k}B^{i}\right)  \psi_{l}$ & $-\lambda_{l}\sin\theta\psi
_{l}^{\dagger}\gamma^{0}\left(  \mathcal{R}_{ik}\Sigma^{k}\tilde{B}%
^{i}\right)  \psi_{l}$\\\hline
\end{tabular}
\caption{EDM, WEDM, MDM and WMDM contibutions to the hamiltonian of the lepton
nonminimal coupling (\ref{Lepton1}).}%
\label{T1a}%
\end{table}
\begin{table}[t]
\centering
\begin{tabular}
[c]{|c|c|c|c|c|c|c|}\hline
Coupling & $\lambda_{l}T$ & $\lambda_{l}T_{ij}$ & $\lambda_{l}\mathcal{T}%
_{jk}$ & $\lambda_{l}R_{00}$ & $\lambda_{l}R_{ij}$ & $\lambda_{l}%
\mathcal{R}_{ij}$\\\hline
P & $+$ & $+$ & $-$ & $-$ & $-$ & $+$\\\hline
C & $+$ & $+$ & $+$ & $+$ & $+$ & $+$\\\hline
T & $+$ & $+$ & $-$ & $-$ & $-$ & $+$\\\hline
\end{tabular}
\caption{Complete classification under $C,P,T$ \ for the $CPT$-even nonminimal
couplings of Table (\ref{T1a}), showing the EDM signature for $\lambda
_{l}\mathcal{T}_{jk},$ $\lambda_{l}R_{00},$ $\lambda_{l}R_{ij},$ and the MDM
behavior for $\lambda_{l}T,\lambda_{l}T_{ij},\lambda_{l}\mathcal{R}_{ij}.$}%
\label{TableCPT2}%
\end{table}

The tau lepton data can be used to constrain the lepton weak EDM and weak MDM
couplings of\ Table (\ref{T1a}). Using the upper bound for the tau lepton WEDM
\cite{Pich}, the element (\ref{LWEDM1}) leads to $\left\vert \lambda_{\tau
}\left(  \sin\theta\right)  \mathcal{T}_{jk}\right\vert <1.2\times
10^{-17}e\cdot cm,$ that is%
\begin{equation}
\left\vert \lambda_{\tau}\mathcal{T}_{jk}\right\vert <1\times10^{-4}\left(
\text{GeV}\right)  ^{-1}, \label{UBLT0i}%
\end{equation}
where we used $\sin\theta=0.48.$ Having in mind the definition $\mathcal{T}%
_{ij}=T_{0a}\epsilon_{aij},$ obviously the tensor has no isotropic component,
$\mathcal{T}_{ii}=0.$ The component $\mathcal{T}_{jk}$ to be constrained
depends on the direction of the electric field. In an apparatus the electric
field points along the z-axis, the components to be restrained are
$\mathcal{T}_{13},\mathcal{T}_{23}.$ The constraining procedure can be applied
on the other WEDM pieces of Eq. (\ref{LWEDM2}), implying the upper bounds:%
\begin{align}
\left\vert \lambda_{\tau}R_{00}\right\vert  &  <1\times10^{-4}\left(
\text{GeV}\right)  ^{-1},\label{tauLT00}\\
\left\vert \lambda_{\tau}R_{ij}\right\vert  &  <1\times10^{-4}\left(
\text{GeV}\right)  ^{-1}. \label{tauLTij}%
\end{align}

Tau WMDM experimental upper bounds \cite{Pich} can also be used to constrain
the tensor components of Table (\ref{T1a}), $\alpha_{w}<1\times10^{-3}$, or
\begin{equation}
\frac{e}{2m_{\tau}}\frac{\alpha_{w}}{\sin2\theta}<3\times10^{-5}%
(\text{GeV})^{-1},
\end{equation}
which is the factor that bounds the WMDM\ coefficients of Lagrangian
(\ref{Lepton1_1}) and (\ref{Lepton1_2}). For the isotropic component, we write
$\sin\theta\left\vert \lambda_{\tau}T\right\vert <3\times10^{-5}($GeV$)^{-1},$
or
\begin{equation}
\left\vert \lambda_{\tau}T\right\vert <6\times10^{-5}(\text{GeV})^{-1}.
\end{equation}
The same holds for $\left\vert \lambda_{\tau}T_{ij}\right\vert $ and
$\left\vert \lambda_{\tau}\mathcal{R}_{ij}\right\vert ,$ as it appears in
Table (\ref{T2b}).

In order to constrain the EDM couplings, we should use the electron EDM
measurements, which represent the smallest EDM limit ever established,
$d_{e}<1.1\times10^{-31}e\cdot m$ \cite{ACME2018}. For the isotropic
component, $\left\vert \lambda_{e}\left(  \cos\theta\right)  R_{00}\right\vert
<1.1\times10^{-31}e\cdot m,$
\begin{equation}
\left\vert \lambda_{e}R_{00}\right\vert <5\times10^{-17}\left(  \text{GeV}%
\right)  ^{-1},
\end{equation}
where we have used $\cos\theta=0.88$, the same holding for the other
components\textbf{ }$\left\vert \lambda_{e}\mathcal{T}_{jk}\right\vert $,
$\left\vert \lambda_{e}R_{ij}\right\vert $.

Concerning the MDM interaction,
\begin{equation}
\mathcal{L}=\bar{\psi}\left[  g\frac{e}{2m_{l}}\sigma^{\mu\nu}F_{\mu\nu
}\right]  \psi,
\end{equation}
we can use the electron data to constrain it. \ The electron's magnetic moment
is $\boldsymbol{\mu}=-g\mu_{B}\boldsymbol{S},$ with $\mu_{B}=e/2m$ being the
Bohr magneton and $g=2(1+a)$ being the gyromagnetic factor, with
$a=\alpha/2\pi\simeq0.00116$ representing the deviation from the usual case,
$g=2$. The magnetic interaction is $H^{\prime}=-\mu_{B}g\left(  \boldsymbol{S}%
\cdot\mathbf{B}\right)  $. Precise measurements reveal that the experimental
imprecision on the electron MDM is at the level of $2.8$\ parts in $10^{13}$
\cite{Gabrielse}$,$\ that is, $\Delta a\leq2.8\times10^{-13}.$ This value
represents the window for new contributions that stem form dimension five
terms, in such a way that $\lambda_{e}T\cos\theta<\mu_{B}\Delta a=2.4\times
10^{-20}($eV$)^{-1},$ implying%
\begin{equation}
\lambda_{e}T<3\times10^{-11}(\text{GeV})^{-1}. \label{MDMway1}%
\end{equation}

We thus observe that the electron EDM data imply better couplings than the
e-MDM data by a factor $10^{6}$, while the tau-WEDM and WMDM imply constraints
of similar magnitude, which is explained by the large tau mass, which yields a
much smaller Bohr magneton. The upper bounds obtained are presented in Tables
(\ref{T2a}) and (\ref{T2b}).

\begin{table}[t]
\centering%
\begin{tabular}
[c]{|c|c|c|}\hline
Coupling & e-EDM & tau-WEDM\\\hline
$\lambda_{l}\mathcal{T}_{jk}$ & $\left\vert \lambda_{e}\mathcal{T}%
_{jk}\right\vert <5\times10^{-17}$ & $\left\vert \lambda_{\tau}\mathcal{T}%
_{jk}\right\vert <1\times10^{-4}$\\\hline
$\lambda_{l}T$ & $-$ & $-$\\\hline
$\lambda_{l}T_{ij}$ & $-$ & $-$\\\hline
$\lambda_{l}R_{ij}$ & $\left\vert \lambda_{e}R_{ij}\right\vert <5\times
10^{-17}$ & $\left\vert \lambda_{\tau}R_{ij}\right\vert <1\times10^{-4}%
$\\\hline
$\lambda_{l}\mathcal{R}_{jk}$ & $-$ & $-$\\\hline
$\lambda_{l}R_{00}$ & $\left\vert \lambda_{e}R_{00}\right\vert <5\times
10^{-17}$ & $\left\vert \lambda_{\tau}R_{00}\right\vert <1\times10^{-4}%
$\\\hline
\end{tabular}
\caption{Upper bounds to EDM and WEDM contibutions to leptons, in $\left(
\text{GeV}\right)  ^{-1}.$}%
\label{T2a}%
\end{table}\begin{table}[t]
\centering%
\begin{tabular}
[c]{|c|c|c|}\hline
Coupling & e-MDM & \ tau-WMDM\\\hline
$\lambda_{l}\mathcal{T}_{jk}$ & $-$ & $-$\\\hline
$\lambda_{l}T$ & $\left\vert \lambda_{e}T\right\vert <3\times10^{-11}$ &
$\left\vert \lambda_{\tau}T\right\vert <6\times10^{-5}$\\\hline
$\lambda_{l}T_{ij}$ & $\left\vert \lambda_{e}T_{ij}\right\vert <3\times
10^{-11}$ & $\left\vert \lambda_{\tau}T_{ij}\right\vert <6\times10^{-5}%
$\\\hline
$\lambda_{l}R_{ij}$ & $-$ & $-$\\\hline
$\lambda_{l}\mathcal{R}_{jk}$ & $\left\vert \lambda_{e}\mathcal{R}%
_{ik}\right\vert <3\times10^{-11}$ & $\left\vert \lambda_{\tau}\mathcal{R}%
_{ik}\right\vert <6\times10^{-5}$\\\hline
$\lambda_{l}R_{00}$ & $-$ & $-$\\\hline
\end{tabular}
\caption{Upper bounds to MDM and WMDM contibutions to leptons, in $\left(
\text{GeV}\right)  ^{-1}.$}%
\label{T2b}%
\end{table}The same kind of analysis holds for the neutrino NM coupling
contained in Lagrangian term (\ref{Neutrino1}), which can be analogously
separated into two pieces,%
\begin{align}
&  \mathcal{L}_{T}^{\nu}=\lambda_{\nu}\bar{\psi}_{\nu}\sigma^{\mu\beta}%
{}\left(  T_{\mu\nu}B_{\text{ \ }\beta}^{\nu}\right)  \psi_{\nu}%
,\label{NNAC}\\
&  \mathcal{L}_{A}^{\nu}=-i\lambda_{\nu}\bar{\psi}_{\nu}\left(  \sigma
^{\mu\beta}\gamma_{5}R_{\mu\nu}B_{\text{ \ }\beta}^{\nu}\right)  \psi_{\nu
}.\label{NAC}%
\end{align}
Due to the similar structure between the lepton and neutrino NM couplings in
Eqs. (\ref{Lepton1}) and (\ref{Neutrino1}), the EDM, WEDM, MDM and WMDM
Lagrangian contributions for neutrinos are, in principle, the same ones of
Table (\ref{T1a}), only by replacing $\psi_{_{l}}\rightarrow\psi_{\nu}$ and
$\bar{\psi}_{_{l}}\rightarrow\bar{\psi}_{\nu}.$ 

\subsection{Rank$-4$ dimension five nonminimal LV electroweak couplings}

In this section, we introduce, directly on the GSW model Lagrangian, the
rank$-4$ dimension five nonminimal LV couplings:%
\begin{align}
\mathcal{L}_{(4)}^{\left(  even\right)  }  &  =\frac{\lambda_{l}}{2}\bar{\psi
}_{_{l}}\left[  \sigma^{\mu\nu}K_{\mu\nu\alpha\beta}B_{\text{ \ }}%
^{\alpha\beta}+i\sigma^{\mu\nu}\gamma_{5}\bar{K}_{\mu\nu\alpha\beta}B_{\text{
\ }}^{\alpha\beta}\right]  \psi_{_{l}}\\
&  +\frac{\lambda_{\nu_{l}}}{2}\bar{\psi}_{\nu_{l}}\left[  \sigma^{\mu\nu
}K_{\mu\nu\alpha\beta}B_{\text{ \ }}^{\alpha\beta}+i\sigma^{\mu\nu}\gamma
_{5}\bar{K}_{\mu\nu\alpha\beta}B_{\text{ \ }}^{\alpha\beta}\right]  \psi
_{\nu_{l}}, \label{LEDM4}%
\end{align}
where the rank-$4$ background tensors $K_{\mu\nu\alpha\beta}$, $\bar{K}%
_{\mu\nu\alpha\beta}$ are antisymmetric in the two pairs:%
\begin{align}
K_{\mu\nu\alpha\beta}  &  =-K_{\nu\mu\alpha\beta},\\
K_{\mu\nu\alpha\beta}  &  =-K_{\mu\nu\beta\alpha}.
\end{align}
Supposing $T_{\nu\beta}=(K)_{\text{ \ }\nu\alpha\beta}^{\alpha}$ and
$R_{\nu\beta}=(\bar{K})_{\text{ \ }\nu\alpha\beta}^{\alpha}$, one can propose
the prescription,
\begin{align}
(K)_{\mu\nu\alpha\beta}  &  =\frac{1}{2}\left(  g_{\mu\alpha}T_{\nu\beta
}-g_{\mu\beta}T_{\nu\alpha}+g_{\nu\beta}T_{\mu\alpha}-g_{\nu\alpha}T_{\mu
\beta}\right)  ,\label{presc}\\
(\bar{K})_{\mu\nu\alpha\beta}  &  =\frac{1}{2}\left(  g_{\mu\alpha}R_{\nu
\beta}-g_{\mu\beta}R_{\nu\alpha}+g_{\nu\beta}R_{\mu\alpha}-g_{\nu\alpha}%
R_{\mu\beta}\right)  ,
\end{align}
where the tensors $T_{\nu\beta},R_{\nu\beta}$ are now symmetric and traceless.
Replacing such a prescription in the lepton sector of the Lagrangian
(\ref{LEDM4}), we obtain:%
\begin{equation}
\mathcal{L}_{(4)l}^{\left(  even\right)  }=\frac{\lambda_{l}}{2}\bar{\psi
}_{_{l}}\left[  \sigma^{\alpha\nu}T_{\nu\beta}B_{\alpha}^{\text{ \ \ }\beta
}+i\sigma^{\alpha\nu}\gamma_{5}R_{\nu\beta}B_{\alpha}^{\text{ \ \ }\beta
}\right]  \psi_{_{l}}.
\end{equation}
These couplings recover the ones involving ranking-2 tensors, already
presented. Thus, if the rank-4 tensor is written as shown in expression
(\ref{presc}), the upper bounds found in the last section hold equivalently
for some components of $(K)_{\mu\nu\alpha\beta}.$ For instance, $T_{00}=-$
$(K)_{i0i0}$ and $T_{ij}=$ $(K)_{0i0j}-(K)_{ninj},$ so that the WEDM upper
limits (\ref{tauLT00}) and (\ref{tauLTij}) are read as:
\begin{align}
\left\vert \lambda_{\tau}(K)_{i0i0}\right\vert  &  <1\times10^{-4}\left(
\text{GeV}\right)  ^{-1},\\
\lambda_{\tau}\left\vert (K)_{0i0j}-(K)_{ninj}\right\vert  &  <1\times
10^{-4}\left(  \text{GeV}\right)  ^{-1}.
\end{align}

\section{Conclusion and final remarks}

We analyzed dimension five LV nonminimal couplings in the EW sector. The
CPT-odd ones are not effective in generating EDM or MDM contributions, both in
the rank-1 and rank$-3$ forms. Such impossibility is confirmed by the
EDM-incompatible signature under C,P and T operators, as shown in Table
(\ref{Tab1}). We also examined CPT-even nonminimal electroweak couplings,
which generate tree level EDM, MDM, WEDM and WMDM contributions. We firstly
have introduced rank$-2$ dimension five nonminimal couplings directly in the
GSW model Lagrangian, using two rank-$2$ background tensors, $T_{\mu\nu}$ and
$R_{\mu\nu},$ as presented in Lagrangians (\ref{Lepton1piece}) and
(\ref{Lepton2piece}). We have identified the coefficients that generate EDM,
MDM, WEDM and WMDM lepton contribution to the Hamiltonian. Then, we used
experimental data of tau lepton to constrain the WEDM\ and WMDM couplings to
the level of $10^{-4}\left(  \text{GeV}\right)  ^{-1},$ and electron MDM and
EDM data to constrain EDM and MDM couplings to the level of $10^{-17}\left(
\text{GeV}\right)  ^{-1}$ and $10^{-11}\left(  \text{GeV}\right)  ^{-1},$
respectively. These upper bounds are shown in Tables (\ref{T2a}) and
(\ref{T2b}). We have also proposed CPT-even nonminimal EW couplings involving
a rank-$4$ background tensor, $K_{\mu\nu\beta\alpha},$ coupled to the $U(1)$
field strength and the leptons' (neutrinos') spinors. Using a suitable
relation, Eq. (\ref{presc}), we showed that some rank-4 couplings become
equivalent to rank-2 couplings. Thus, the rank-$4$ nonminimal couplings that
generate EDM, MDM, WEDM and WMDM are bounded to the same level of constraining
presented in Tables (\ref{T2a}) and (\ref{T2b}).

\section{Acknowledgements}

\begin{acknowledgments}
The authors are grateful to CNPq, CAPES and FAPEMA (Brazilian research
agencies) for invaluable financial support.
\end{acknowledgments}

\end{document}